\documentclass{aastex631}
\usepackage{amsmath}
\usepackage{epstopdf}
\usepackage{float}
\usepackage{graphicx}
\usepackage{listings}
\usepackage{rotating}
\usepackage{ulem}

\newcommand{\lapprox} {\, \lower3pt\hbox{$\sim$}\llap{\raise2pt\hbox{$<$}}\,}
\newcommand{\gapprox} {\, \lower3pt\hbox{$\sim$}\llap{\raise2pt\hbox{$>$}}\,}

\begin{document}


\title{Temperature and Differential Emission Measure Profiles in Turbulent Solar Active Region Loops}

\author[0000-0001-8720-0723]{A. Gordon Emslie}
\affiliation{Department of Physics \& Astronomy, Western Kentucky University, Bowling Green, KY 42101, USA}

\author[0000-0002-3300-6041]{Stephen J. Bradshaw}
\affiliation{Department of Physics \& Astronomy, Rice University, Houston, TX 77005, USA}

\begin{abstract}

We examine the temperature structure of static coronal active region loops in regimes where thermal conductive transport is driven by Coulomb collisions, by turbulent scattering, or by a combination of the two. (In the last case collisional scattering dominates the heat transport at lower levels in the loop where temperatures are low and densities are high, while turbulent scattering dominates the heat transport at higher temperatures/lower densities.) Temperature profiles and their corresponding differential emission measure distributions are calculated and compared to observations, and earlier scaling laws relating the loop apex temperature and volumetric heating rate to the loop length and pressure are revisited. Results reveal very substantial changes, compared to the wholly collision-dominated case, to both the loop scaling laws and the temperature/density profiles along the loop.  They also show that the well-known excess of differential emission measure at relatively low temperatures in the loop may be a consequence of for by the flatter temperature gradients (and so increased amount of material within a specified temperature range) that results from the predominance of turbulent scattering in the upper regions of the loop.

\end{abstract}

\section{Introduction}\label{introduction}

On a variety of grounds, both theoretical and observational, it is becoming increasingly clear that turbulence plays an important role in determining the structure of the active solar atmosphere. Turbulent flows are a natural consequence of the high Reynolds number in the solar corona \citep[e.g.,][]{2014masu.book.....P}, and some form of turbulence on the micro-scale is necessary to create a plasma resistivity sufficiently large to rapidly release energy over large spatial scales during the impulsive phase of solar flares \citep[e.g.,][]{1971ApJ...169..379C}. Magnetohydrodynamic (MHD) turbulence (i.e., stochastic motions within the magnetized plasma) is a leading candidate for transferring energy released in the primary (magnetic reconnection) energy release site into the production of the energetic particles that are the hallmark of the impulsive phase of a solar flare \citep[e.g.,][]{1993ApJ...418..912L, 1996ApJ...461..445M,2012SSRv..173..535P}; this transfer of energy generally proceeds via the turbulent cascade of energy to progressively smaller scales, eventually dissipating at the particle level. EUV and soft X ray spectral lines observed during flares frequently have a width that is significantly in excess of the thermal Doppler width \citep{1982SoPh...78..107A,1990A&A...236L...9A,1992str..book.....M,2010A&A...521A..51P,2015ApJ...799L..12D} and it is generally accepted that such broadening is strong evidence for the presence of micro-turbulent \citep{1982SoPh...78..107A,1992str..book.....M,2010A&A...521A..51P} and/or macro-turbulent \citep{1992str..book.....M,2010A&A...521A..51P} flows. Cotemporaneous observations using multiple instruments have shown \citep{2017PhRvL.118o5101K} that hydrodynamic turbulence is not only located near the site of flare primary energy release, but also has a sufficient energy content to play a major role in the transfer of energy from the primary magnetic reconnection into other manifestations of the flare, such as accelerated nonthermal particles.

Even so-called ``static'' active region loops are currently believed to be heated by a canonical \citep{1988ApJ...330..474P} process involving the creation of many small current sheets in the corona, with a repeat time that is short relative to the cooling time; such a coronal heating scenario requires some anomalous microscale process to generate a large enough plasma resistivity to dissipate energy on spatial scales of $\sim$1~Mm. \cite{2021NatAs...5..103A} have detected the bidirectional jets (``nanojets'') predicted to be associated with reconnection in small-scale current sheets (nanoflares), and \cite{2021NatAs...5..237B} found excess line broadenings in the vicinity of small-scale reconnection events, with a dependence on ion species that was consistent with ion-cyclotron turbulence. Even if heating does not originate from small-scale reconnection events but instead is purely wave-related (e.g., via oppositely propagating Alfv\'en waves) then turbulence is still required to dissipate energy at the interacting wave fronts.

As noted by several authors \citep[e.g.,][]{2014ApJ...795..171S,2013ApJ...778...68R}, the observed cooling times of active region and post-flare loops are significantly greater than those predicted by conventional energy transport models that involve collision-dominated \citep{1962pfig.book.....S} conduction. This has been taken as evidence for continuous ongoing small-scale heating of such loops, but also \citep[as discussed by][]{2016ApJ...824...78B} may be indicative of a suppressed thermal conductive flux, as might be appropriate to an environment in which thermal conduction is driven by turbulent, rather than collisional, scattering.

In summary, the presence of turbulence is not necessarily restricted to flaring regions; it is also likely that a significant level of turbulence is present in steady-state active region loops. This presents a compelling rationale to consider the effect of turbulence on the thermodynamic structure of such loops, which is the purpose of the present work.

Considerable progress in our understanding of the structure of coronal loops has been made since the seminal work of \cite{1978ApJ...220..643R}, based on {\it Skylab} ATM data. In particular, a considerable body of work has been carried out to determine the degree to which coronal loops can be considered as a single flux tube or as a tangled set of thermally isolated ``strands.'' \cite{2008ApJ...686L.131W} presented observations from the Hinode EUV Imaging Spectrometer \citep[EIS;][]{2007SoPh..243...19C} of localized regions near the top of several coronal loop structures. Both delta-function (i.e., isothermal) and Gaussian fits to the differential emission measure ($DEM$) profile were attempted, and it was concluded that an isothermal fit was in general not justified, and hence that these observations ``lend support to the nonequilibrium, multithread models.''  It should be noted that these $DEM$ profiles reflect temperature variations both parallel and perpendicular to the guiding magnetic field (although the selection of pixels near the apex of the loop structures presumably limited the former), and that the use of a Gaussian profile for the $DEM$ distribution did not allow for the identification of higher temperature (e.g., soft X-ray) components in the loop.

\cite{2001ApJ...556..896S} used results from the SoHO Coronal Diagnostics Spectrometer \citep[CDS;][]{1995SoPh..162..233H} and Yohkoh Soft X-Ray Telescope \citep[SXT;][]{2007SoPh..243...63G} to construct the $DEM$ by using a manual iterative method involving forward-fitting a more general form than the Gaussian profile used by \cite{2008ApJ...686L.131W}.  They concluded that ``the temperature distributions are clearly inconsistent with isothermal plasma along either the line of sight or the length of the loop.'' The $DEM$ distributions obtained were peaked at a value $\log T \simeq 6.25$, with a half-width of $\Delta \log T \simeq \pm 0.25$. Further observations of active region loops using the Hinode EIS was carried out by \cite{2009ApJ...694.1256T}, who used the ``EM Loci'' method (essentially, a map of intensity divided by the product of the species abundance and line emissivity function $G(T)$) to delineate loop structures. They concluded that the overall structure of coronal loops becomes less defined (``fuzzier'') at higher temperatures and that the loops are ``almost isothermal along the line of sight.'' However, they also concluded that the filling factors are significantly less than unity, consistent with a multi-strand model.

\cite{2005ApJ...627L..81S} used observations of a coronal loop on the solar limb, observed by the SoHO CDS, to deduce that ``the plasma was multithermal, both along the length of the loop and along the line of sight.''  Noting that other authors had obtained very different results, using data from different instruments (such as the Transition Region and Coronal Explorer \citep[TRACE;][]{1999SoPh..187..229H}, they suggested that ``a variety of temperature structures may be present within loops.'' A later paper \citep{2007ApJ...658L.119S} used simultaneous CDS observations of two loops located side-by-side on the solar disk, with all pixels in both loops visible within the CDS slit. Both forward-fitting EM Loci, and an automated inversion analysis that represents the $DEM$ profile as a series of spline knots \citep{2005ApJS..157..147W} showed that one loop was consistent with a delta-function $DEM$ (i.e., was indistinguishable from isothermal), while the other loop required a broad $DEM$, and so not consistent with an isothermal plasma. \cite{2008ApJ...684L.115S} used observations taken on 2007 May~1 using the Hinode EIS to show that the observed intensities were consistent both with a single-peaked $DEM$ (i.e., isothermal) and a double-peaked $DEM$ (i.e., a sum of two nearly isothermal loops). It was noted, however, that the broadening of each of these components was such that they could not ``simply represent two isothermal strands of the EIS loop or two isothermal loops along the line of sight,'' but ``could, however, represent either two dominant ensembles of strands for the observed EIS loop or the dominant ensemble of strands for two individual loops along the line of sight.'' \cite{2011ApJ...740....2W} also found that the $DEM$ profile was ``broad and peaked around 3~MK,'' but also \citep{2012ApJ...746L..17W} noted that a ``blind spot'' exists in temperature-emission measure space for the combined Hinode EIS and XRT observations they used, so that this data set is insensitive to plasma with temperatures greater than $\simeq 6$~MK.

In \cite{2009ApJ...691..503S}, observations of three different loops were carried out under a Joint Observing Program involving both the TRACE and CDS instruments, which had been noted in \cite{2005ApJ...627L..81S} to have yielded contradictory results.  The introduction to that paper succinctly summarized the isothermal/multithermal dilemma: ``For example, the loops analyzed by \cite{2003A&A...406L...5D} and \cite{2003A&A...406.1089D} were isothermal along the line of sight and had a temperature gradient along their length, but the one analyzed by \cite{2002A&A...383..661B} was isothermal in both directions. The loop analyzed by \cite{2001ApJ...556..896S} and \cite{2006ApJ...636L..49S} was multithermal along the line of sight, with a temperature distribution that increased as a function of loop height.''  It was found in \cite{2009ApJ...691..503S} that data from both instruments supported an isothermal model for one of the loops and a multithermal model for another, but that results for the other two loops were not as clear, due to complicating observational factors such as overlapping structures in the field of view. \cite{2009ApJ...691..503S} concluded that the answer to the question ``Are Coronal Loops Isothermal or Multithermal?'' was a rather perplexing ``yes.''

\cite{2010ApJ...723.1180S} \citep[see also][]{2011ApJ...738..146S} applied Monte Carlo based, iterative forward fitting $DEM$ algorithms to data from the Hinode XRT and EIS instruments, and found that the observations were consistent with a $DEM$ profile with a peak at $\log T \simeq 6.5$ and a fairly narrow half-width $\Delta \log T \simeq 0.25$ \citep[similar to the results obtained by][]{2001ApJ...556..896S}. They concluded that ``at least some loops are not consistent with isothermal plasma, and therefore cannot be modeled with a single flux tube and must be multi-stranded.''

The launch of the Solar Dynamics Observatory and its high-spatial-resolution Atmospheric Imaging Assembly \citep[AIA;][]{2012SoPh..275...17L} with several different wavelength filters, representing a wide variety of line formation temperatures, opened up a new era in the interpretation of loop structures. \cite{2010ApJ...725L..34S} reported their analysis of a loop observed by AIA on 2010 August~3. Figure~4 of that paper shows the ratios of  model-to-observational fluxes in six different AIA spectral channels, and shows convincingly that while an extended, but relatively narrow ($\log T \simeq 6.35; \Delta \log T \simeq 0.25$) $DEM$ profile could straightforwardly account for the observations in each AIA channel, a delta-function $DEM$ profile (i.e., isothermal plasma) could not, with model-to-observation ratios ranging from less than unity to almost ten in different channels. The width of the $DEM$ profile obtained was consistent with the previous results of \cite{2001ApJ...556..896S} and \cite{2010ApJ...723.1180S}. \cite{2011ApJ...731...49S} then focused on a dozen relatively cool loops that are prominent in the 171~\AA\ channel of AIA, which has a peak response temperature of $\log T \simeq 5.8$. They found that one-third of the loops observed had narrow temperature distributions, consistent with isothermal plasma, while other loops had $DEM$ distributions with centroids around $\log T \simeq 6.0 - 6.2$ and half-widths $\Delta \log T$ up to $\simeq 0.4$, somewhat broader than previously obtained by \cite{2001ApJ...556..896S,2010ApJ...725L..34S,2010ApJ...723.1180S}. This analysis was extended to loops prominent in the 211~\AA\ channel by \cite{2011ApJ...739...33S}, which has a higher peak response temperature ($\log T \simeq 6.3$). Results were inconclusive, however, because one of the AIA channels (at 131~\AA\ ) contains Fe VII emission not only from relatively cool ($\log T \simeq 5.7$) material, but also from Fe~XX and Fe~XXIII lines, which are formed at much higher temperatures ($\log T \simeq 7.2$). Further, analysis excluding the problematic 131~\AA\ data proved to be inadequate to usefully constrain the $DEM$ profile. Using improved atomic data from CHIANTI 7.1, \citep{2013ApJ...770...14S} were able to construct $DEM$ profiles for the same set of loops; these profiles (their Figure~4) were roughly Gaussian in $(\log T, \log DEM)$ space, with peaks around $\log T \simeq 6.0$ and standard deviations $\Delta \log T \simeq 0.3$.  Similar results (their Figure~8) were reported by \cite{2013ApJ...764...53S} using Hinode EIS and XRT data. Pursuing this further, \cite{2014ApJ...795..171S} used data from all three instruments (Hinode XRT, Hinode EIS, and SDO AIA) to show that cooler loops tend to have narrower $DEM$ widths.

The essential results of the above papers are summarized in the ``Coronal Loop Inventory Project'' papers of \cite{2015ApJ...813...71S} and \cite{2016ApJ...831..199S}. Basically, the  $DEM$ of coronal loops, while it can be consistent with an ``isothermal'' delta-function, may also be ``multithermal.'' Thus, in general, a loop must be considered as a collection of individual ``strands,'' each with its own one-dimensional field-aligned temperature profile. Using the sub-orbital rocket-borne Hi-C \citep{2014SoPh..289.4393K} instrument, \cite{2013Natur.493..501C} have discovered direct observational evidence for the presence of such finely braided loop ``strands.''

Given this observational background, it must nevertheless be noted that even for the ``multithermal'' case of a heterogeneous cross-field temperature structure, or more generally, when the three-dimensional nature of an ensemble of loops is taken into account \citep{1999ApJ...515..842A,2016ApJ...821...63B,2019ApJ...880...56B,2021ApJ...919..132B}, the temperature and density structure along the strand is still ubiquitously modeled by a one-dimensional energy balance model, as originally presented by \cite{1978ApJ...220..643R}. We therefore here consider the influence of turbulence on the temperature (and $DEM$) structure of loops through the modeling of individual one-dimensional strands; convolution of these results with a suitable cross-field representation of density and peak temperature may then be used to construct model ``loops.''  Of particular note in the context of the present work are the works by \cite{2007A&A...469..347B} and \cite{2019ApJ...883...20G}, who modeled the role of turbulence in the {\it heating} of coronal loops, the latter by invoking turbulence generated by the Kelvin-Helmholtz instability associated with the interaction of multiple strands within a loop.

The presence of microturbulence has a very significant effect on the plasma transport coefficients and thus on the transport of particles and their associated momentum and energy fluxes \citep{2016ApJ...824...78B}. Modeling using the enthalpy-based thermal evolution of loops  \citep[EBTEL;][]{2008ApJ...682.1351K,2010ApJ...710L..39B,2012ApJ...752..161C,2012ApJ...758....5C} code has shown \citep{2018ApJ...852..127B} that suppression of heat conduction by turbulence can help explain the anomalously long cooling times observed \citep{1980sfsl.work..341M,2013ApJ...778...68R} in post-flare loops, reducing (but not necessarily eliminating) the level of heating necessary in the post-impulsive phase of a flare. Turbulence also affects the parallel thermoelectric transport coefficients $(\kappa, \alpha, \beta, \sigma)$ that appear in the relations \citep{2016ApJ...824...78B} connecting the heat flux $q$ and electrical current density $j$ to the parallel temperature gradient $dT/dz$ and the local parallel electric field $E_\parallel$:

\begin{equation}\label{thermoelectric-coefficients}
\begin{pmatrix}
q \\
j
\end{pmatrix}
=
\begin{pmatrix}
-\kappa & -\alpha \\
\beta & \sigma
\end{pmatrix}
\begin{pmatrix}
\frac{dT}{dz} \\
E_\parallel
\end{pmatrix} \,\,\, .
\end{equation}
In general, all four coefficients scale approximately as $1/R$, where the suppression factor $R$ is the ratio of the collisional to turbulent mean free paths for electrons moving at the local thermal speed.

As explained by \cite{2019ApJ...880...80B}, any turbulence-related reduction in the thermal conduction coefficient $\kappa$ acts to fundamentally change the scaling laws \citep{1978ApJ...220..643R} that relate the temperature and volumetric heating rate in a static active region coronal loop to its density and length. For example, if the turbulent scattering mean free path $\lambda_T$ is taken to be a constant, independent of both density and velocity, the resulting much weaker temperature dependence of the thermal conduction coefficient, compared to that associated with collisional scattering \citep{1962pfig.book.....S}, leads to a temperature gradient that is more uniform over the loop. Accordingly, it is now the lower temperature regions, with their correspondingly higher densities, that dominate the overall energy balance, so that the loop scaling laws now involve not just the peak temperature $T_{\rm max}$ but also the loop base temperature $T_o$. The effects of turbulence have an even more profound effect on the scaling laws for {\it dynamic} loops, effectively limiting \citep{2020ApJ...904..141B} the speed of any flows generated.

Such a reduction in the parallel conductive heat transport coefficient $\kappa$ would be expected to result in significant changes to the temperature/height and density/height profiles within the loop, that in turn determine the differential emission measure vs. temperature profile and so the emissivity properties of the loop plasma. In this work we therefore explore beyond the scaling law results of \citep{2019ApJ...880...80B} to consider (Section~\ref{analysis}) the temperature profiles of the loops when the thermal conduction term in the steady-state energy equation is dominated by turbulent, rather than collisional \citep[see][]{2010ApJ...714.1290M}, transport. (This more precise analysis also enables us to refine the approximate scaling laws obtained by \cite{2019ApJ...880...80B}.)  Then, recognizing that the collisional mean free path is both temperature- and density-dependent, and hence that the relative importance of collisional and turbulent transport varies with position along an active region coronal loop, we then consider a hybrid model formed by juxtaposing a turbulence-dominated model at high temperatures with a collision-dominated model at low temperatures. We further pursue such hybrid modeling by considering a numerical treatment in which the ratio of collisional to turbulent scattering mean free paths evolves continuously with position along the loop.

Our results show that including the effects of turbulent scattering (whether throughout the entire loop or only in the high-temperature regions) produces flatter temperature gradients, and so larger differential emission measures (Section~\ref{dem-profiles-section}), than for a loop in which collisional transport prevails throughout. Comparison of these model predictions with observations indicates much better agreement than for collision-dominated models, and hence that turbulence indeed plays a significant role in determining the thermodynamic structure of active region loops. Implications for a variety of other studies, including the response of active region loops to impulsive energy deposition during solar flares, are then examined in Section~\ref{conclusions}.

\section{Derivation of the Scaling Laws and Temperature Profiles}
\label{analysis}

We begin by reviewing and, where appropriate, generalizing the work of \cite{2010ApJ...714.1290M} on the temperature structure of active region loop ``strands'' in steady-state energy balance between heat input, radiative losses, and conductive redistribution.  The energy equation \citep[cf. Equations~(1) and~(2) of][]{2010ApJ...714.1290M} is

\begin{equation}\label{energy-equation-general}
\frac{d}{dz} \left ( \kappa_o \, T^\delta \, \frac{dT}{dz} \right ) + H \, P^\beta \, T^\alpha - \left(\frac{P}{2 k_B}\right)^2 \, \chi_o \, T^{-(2 + \gamma)} = 0 \,\,\, ,
\end{equation}
where $z$ is the distance along the loop (measured upward from the chromospheric footpoint toward the loop apex situated at a distance $z=L$), $T$ (K) is the electron temperature, and $P = 2 n k_B T$ (dyne~cm$^{-2}$) is the loop pressure (assumed uniform). In addition, $k_B = 1.38 \times 10^{-16}$~erg~K$^{-1}$ is Boltzmann's constant, the dimensionless parameters $\alpha$ and $\beta$ characterize the temperature and pressure dependence of the loop heating function (the magnitude of which is characterized by the parameter $H$), and $\chi_o$ and $\gamma$ define respectively the magnitude and temperature dependence of the optically thin radiative loss function $\Phi(T) = \chi_o \, T^{-\gamma}$ \citep[for the temperature range under consideration, $\chi_o = 1.6 \times 10^{-19}$~erg~cm$^3$~s$^{-1}$~K$^{1/2}$ and $\gamma = 1/2$; e.g.,][]{1969ApJ...157.1157C}. In Equation~(\ref{energy-equation-general}), we have generalized the conduction term, both in its magnitude $\kappa_o$ and in its temperature dependence $T^\delta$.

\subsection{Collision-Dominated Conduction}\label{collision-dominated}

For collision-dominated conduction, the pertinent mean free path is that appropriate to Coulomb collisions, viz.

\begin{equation}\label{mean-free-path-collisional}
\lambda_C = \frac{(2 k_B)^2}{2 \pi e^4 \Lambda} \, \frac{T^2}{n} \,\,\, ,
\end{equation}
where $e= 4.8 \times 10^{-10}$~esu is the electronic charge and $\Lambda \simeq 20$ is the Coulomb logarithm \citep{1962pfig.book.....S}.  The coefficient of the temperature gradient in the energy equation~(\ref{energy-equation-general}) is thus

\begin{equation}\label{conductive-parameters-coll}
2 n k_B \sqrt{\frac{2 k_B T}{m_e}} \, \lambda_C =
\frac{2^{5/2} \, k_B^{7/2}}{\pi m_e^{1/2} e^4 \Lambda} \, T^{5/2} \equiv \kappa_{oc} \, T^{\delta_C} \,\,\, ,
\end{equation}
where $m_e = 9.1 \times 10^{-28}$~g is the electron mass and we have made the identifications

\begin{equation}\label{conductive-parameters-collisional}
\kappa_{oc} = \frac{2^{5/2} \, k_B^{7/2}}{\pi m_e^{1/2} e^4 \Lambda} \, ; \qquad \delta_C = \frac{5}{2} \,\,\, .
\end{equation}

As shown by \cite{2010ApJ...714.1290M}, the loop temperature profiles, as least for collision-dominated conduction, are insensitive to the values of the parameters $\alpha$ and $\beta$ that appear in Equation~(\ref{energy-equation-general}). We shall therefore take both $\alpha$ and $\beta$ to be zero, so that the volumetric heating is uniform. (Generalizing the analysis below to other values of $\alpha$ and $\beta$ is straightforward.) Introducing the dimensionless variables

\begin{equation}\label{dimensionless-vars}
\eta = \left ( \frac{T}{T_{\rm max}} \right )^{7/2} \,; \qquad x = \frac{z}{L} \,\,\, ,
\end{equation}
where $T_{\rm max}$ is the temperature at the loop apex, and the dimensionless parameters

\begin{equation}\label{dimensionless-parameters}
\epsilon = \frac{2 \, \kappa_{oc} \, (2 k_B)^2 \, T_{\rm max}^{6}}{7 \, \chi_o \, P^2 \, L^2} \, ; \qquad \xi = \frac{H \, \, (2 k_B)^{2} \, T_{\rm max}^{5/2}}{\chi_o \, P^{2}}
\end{equation}
\citep[cf. Equations (3) through (8) of][but with $P=2 n k_B T$ replacing Martens' $P_o = nT$]{2010ApJ...714.1290M}, the energy equation can be written in the form

\begin{equation}\label{eta-double-prime}
\epsilon \, \eta^{\prime \prime} = \eta^{-5/7} - \xi  \,\,\, .
\end{equation}
This has a first integral

\begin{equation}\label{first-integral}
\frac{\epsilon}{2} \, {\eta^{\prime}}^2 = \frac{7}{2} \, \eta^{2/7} - \xi \,  \eta \,\,\, ,
\end{equation}
where, following the argument of \cite{2010ApJ...714.1290M}, we have used the boundary condition that the temperature gradient vanish at the loop base, i.e., $\eta^\prime = 0$ when $\eta=0$.

If collision-dominated conduction holds throughout the loop \citep[as tacitly assumed by][]{2010ApJ...714.1290M}, then, by symmetry at the loop apex, we can also set $\eta^\prime = 0$ when $\eta=1$.  This implies that

\begin{equation}\label{xi-collisional}
\xi = 7/2 \,\,\, ,
\end{equation}
i.e., using Equation~(\ref{dimensionless-parameters}),

\begin{equation}\label{scaling-law-2-Martens-final}
H = \frac{7 \, \chi_o}{8 \, k_B^2} \, \frac{P^{2}}{T_{\rm max}^{5/2}} \,\,\, .
\end{equation}
Further, with $\xi=7/2$, Equation~(\ref{first-integral}) becomes

\begin{equation}\label{first-integral-case-1}
\frac{\epsilon}{2} \, {\eta^{\prime}}^2 = \frac{7}{2} \left( \eta^{2/7} - \eta \right) \,\,\, ,
\end{equation}
which can be directly integrated to obtain $\eta$ as an implicit function of $x$:

\begin{equation}\label{solution-case-1}
\int_0^\eta \frac{d \eta}{\sqrt{\eta^{2/7} - \eta}} = \sqrt{\frac{7}{\epsilon}} \, x \,\,\, .
\end{equation}
Applying Equation~(\ref{solution-case-1}) at the loop apex ($x=1, \eta=1$) gives

\begin{equation}\label{epsilon-case-1}
\epsilon = 7 \, \left [ \int_0^1 \frac{d \eta}{\sqrt{\eta^{2/7} - \eta}} \right ]^{-2} \,\,\, .
\end{equation}
The integral in this expression can be evaluated analytically:

\begin{equation}
\int_0^1 \frac{d \eta}{\sqrt{\eta^{2/7} - \eta}} = \int_0^1 \eta^{-1/7} \, \left ( 1 - \eta^{5/7} \right )^{-1/2}  d \eta  = \frac{7}{5} \, \int_0^1 x^{1/5} \, \left ( 1 - x \right )^{-1/2}  dx = \frac{7}{5} \, B \left ( \frac{6}{5}, \frac{1}{2} \right ) \simeq 1.4 \times 1.79 \simeq 2.51 \,\,\, ,
\end{equation}
where $B(a,b)$ is the beta function. Using the definition of $\epsilon$ from Equation~(\ref{dimensionless-parameters}), Equation~(\ref{epsilon-case-1}) becomes

\begin{equation}\label{scaling-law-1-Martens}
\frac{2 \, \kappa_{oc} \, (2 k_B)^2 \, T_{\rm max}^{6}}{7 \, \chi_o \, P^2 \, L^2} = \frac{25}{7} \, \left [ B \left ( \frac{6}{5}, \frac{1}{2} \right ) \right ]^{-2} \,\,\,
\end{equation}
\cite[cf. Equation~(23) of][]{2010ApJ...714.1290M}; i.e.,

\begin{equation}\label{scaling-law-1-Martens-final}
T_{\rm max} = \left ( \frac{25 \, \chi_o}{8 \, \kappa_{oc} \, k_B^2} \right )^{1/6} \, \,  \left [ B \left ( \frac{6}{5}, \frac{1}{2} \right ) \right ]^{-1/3} \, (PL)^{1/3} \simeq 1.3 \times 10^3 (PL)^{1/3} \,\,\, ,
\end{equation}
which is the so-called ``first scaling law'' (Equation~(4.3) of \cite{1978ApJ...220..643R}; cf. Equation~(31) of \cite{2019ApJ...880...80B}). Substituting this expression for $T_{\rm max}$ into Equation~(\ref{scaling-law-2-Martens-final}) then yields the ``second scaling law'':

\begin{equation}\label{scaling-law-2-Martens-final-1}
H = \frac{7 \, \chi_o}{8 \, k_B^2} \, \left ( \frac{8 \, \kappa_{oc} \, k_B^2}{25 \, \chi_o} \, \right )^{5/12} \, \left [ B \left ( \frac{6}{5}, \frac{1}{2} \right ) \right ]^{5/6} \, P^{7/6} \, L^{-5/6} \simeq 1.2 \times 10^5 \, P^{7/6} \, L^{-5/6} \,\,\,;
\end{equation}
cf. Equation~(4.4) of \cite{1978ApJ...220..643R} and Equation~(32) of \cite{2019ApJ...880...80B}.

\subsection{Turbulence-Dominated Conduction}\label{turb-dominated}

In a turbulent environment, heat transport by thermal conduction is now controlled by turbulent scattering, and the collisional mean free path $\lambda_C \sim T^2/n$ is replaced by a turbulence mean free path $\lambda_T$. Following \cite{2019ApJ...880...80B}, we here take $\lambda_T$ to be a constant, independent of both density and temperature \cite[the results below can straightforwardly be generalized to other forms of $\lambda_T$; cf.][]{2018ApJ...865...67E}. Correspondingly \citep{2019ApJ...880...80B} the coefficient of the temperature gradient appearing in Equation~(\ref{energy-equation-general}) is

\begin{equation}\label{conductive-parameters-turb}
2 n k_B \sqrt{\frac{2 k_B T}{m_e}} \, \lambda_T =  \frac{(2 k_B)^{1/2} \, \lambda_T \, P}{m_e^{1/2}} \, T^{-1/2} \equiv \kappa_{oT} \, T^{\delta_T} \,\,\, ,
\end{equation}
where we have made the identifications

\begin{equation}\label{conductive-parameters-turbulent}
\kappa_{oT} = \frac{(2 k_B)^{1/2} \, \lambda_T \, P }{m_e^{1/2}}  \, ; \qquad \delta_T = -\frac{1}{2} \,\,\, .
\end{equation}
From Equations~(\ref{conductive-parameters-collisional}) and`(\ref{conductive-parameters-turbulent}), we find the ratio and difference

\begin{equation}\label{kappa-ratio}
\frac{\kappa_{oT}}{\kappa_{oc}} = \frac{\pi e^4 \Lambda \, \lambda_T \, P}{4 k_B^3} \equiv \frac{\lambda_T \, P}{c_R} \, ; \qquad \delta_C - \delta_T = 3\,\,\, ,
\end{equation}
where $c_R \equiv 4 k_B^3/\pi e^4 \Lambda \simeq 3.15 \times 10^{-12}$~erg~cm$^{-2}$~K$^{-3}$  \cite[Equation~(11) of][]{2019ApJ...880...80B}. While the change in the magnitude of $\kappa_o$ is significant, we shall see below that the difference in temperature exponents $\delta_C - \delta_T = 3$ (corresponding to the 3 powers of $T$ associated with the change in the mean free path from $\lambda_C \propto T^2/n \propto T^3/P$ to a quantity $\lambda_T$ that is independent of $T$) is even more significant.

Inserting the quantities $\kappa_{oT}$ and $\delta_T$ in the conduction term, the fundamental energy equation~(\ref{energy-equation-general}) becomes

\begin{equation}\label{energy-equation-turbulent}
\frac{d}{dz} \left ( \kappa_{oT} \, T^{-1/2} \, \frac{dT}{dz} \right ) + H - \left(\frac{P}{2 k_B}\right)^2 \, \chi_o \, T^{-5/2} = 0 \,\,\, .
\end{equation}
If we now introduce the dimensionless parameter and variable

\begin{equation}\label{dimensionless-parameters-turbulent}
\epsilon_T = \frac{2 \, \kappa_{oT} \, (2 k_B)^2 \, T_{\rm max}^{3}}{\chi_o \, P^2 \, L^2} \, ; \qquad \zeta = \left ( \frac{T}{T_{\rm max}} \right )^{1/2} \equiv \eta^{1/7}  \,\,\, ,
\end{equation}
Equation~(\ref{energy-equation-turbulent}) becomes

\begin{equation}\label{eta-double-prime-turbulent}
\epsilon_T \, \zeta^{\prime \prime} = \zeta^{-5} - \xi  \,\,\, ,
\end{equation}
where the heating function parameter $\xi$ is the same as before (Equation~(\ref{dimensionless-parameters})). Applying the symmetrical boundary condition $\zeta^\prime = 0$ at the loop apex $\zeta=1$, we find the first integral

\begin{equation}\label{first-integral-turbulent}
\frac{\epsilon_T}{2} \, {\zeta^{\prime}}^2 = \frac{1}{4} \left ( 1 - \frac{1}{\zeta^4} \right ) + \xi \, (1 -  \zeta ) \,\,\, ,
\end{equation}
which, similar to the collisional case~(\ref{solution-case-1}), can be directly integrated to yield $\zeta$ as an implicit function of $x$:

\begin{equation}\label{solution-upper}
\int_{\zeta_o}^{\zeta} \frac{d \zeta}{\sqrt{ \frac{1}{4} \, \left( 1 - \frac{1}{\zeta^4} \right) + \xi \,  \left( 1 - \zeta \right) }} = \sqrt{\frac{2}{\epsilon_T}} \, (x - \ell)  \,\,\, .
\end{equation}
It is important to note that we have here applied a lower boundary condition $\zeta = \zeta_o$ at $x=\ell$; extending the turbulent solution to $\zeta=0$ is not possible\footnote{In particular the boundary condition $\zeta=\zeta^\prime=0$ at $x=0$ used in the collisional case clearly cannot be applied, as is evident from Equation~(\ref{first-integral-turbulent}). In the notation of \cite{2010ApJ...714.1290M} (cf. his Equation (8)), the value of $\mu$ appropriate to turbulent conduction is $\mu=-5$, rather than the $\mu=-5/7$ appropriate to collisional conduction.  Thus, although there are no singularities in the collisional solution -- a fact explicitly noted by \cite{2010ApJ...714.1290M} after his Equation~(13) --  such singularities {\it do} appear in the case of turbulence-dominated conduction.  An alternative way of looking at this is that Martens' Equation~(11) gives $\xi = 1/(\mu+1) = 7/2$ for collisional conduction, but an impossible (negative) value $\xi=-1/4$ for turbulent conduction.}.  Applying Equation~(\ref{solution-upper}) at the loop apex ($x=1, \zeta = \eta=1$) gives

\begin{equation}\label{solution-upper-boundary}
\int_{\zeta_o}^1 \frac{d \zeta}{\sqrt{ \frac{1}{4} \, \left( 1 - \frac{1}{\zeta^4} \right) + \xi \,  \left( 1 - \zeta \right) }} = \sqrt{\frac{2}{\epsilon_T}} \, (1 - \ell)  \,\,\, ,
\end{equation}
to be compared with Equation~(\ref{epsilon-case-1}) for the collisional case.

\begin{figure}[pht]
\centering
\includegraphics[width=0.6\linewidth]{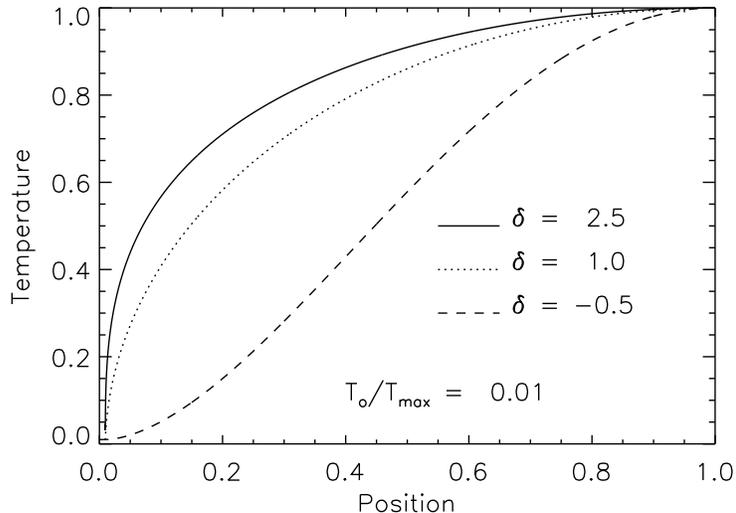}
\caption{\label{t-profiles} Normalized temperature profiles for different values of $\delta$, the temperature index of the thermal conduction coefficient. The case $\delta = 2.5$ corresponds to conduction dominated by collisional transport, while the case $\delta = -0.5$ corresponds to turbulent transport (Equations~(\ref{conductive-parameters-collisional}) and~(\ref{conductive-parameters-turbulent}), respectively.) The case $\delta=1.0$ is also shown to illustrate how the profiles depend on the value of $\delta$.  The turbulent ($\delta = -0.5$) solution applies only down to a lower boundary temperature $T_o$ (see remarks after Equation~(\ref{solution-upper})), here taken to be 0.01 $\times$ the peak loop temperature $T_{\rm max}$.}
\end{figure}

\subsection{Temperature Profiles in Collisional and Turbulent Environments}\label{t-profiles-subsection}

Temperature profiles for both the collisional and turbulent cases can be obtained from Equations~(\ref{solution-case-1}) and~(\ref{solution-upper}), respectively.  (In the latter case a value for the lower temperature $T_o$ must be specified, and from this the value of $\xi$ to be used in Equation~(\ref{solution-upper}) then follows from Equation~(\ref{solution-upper-boundary}).) Figure~\ref{t-profiles} shows the form of the normalized temperature $(T/T_{\rm max}) \equiv \eta^{2/7} \equiv \zeta^2$ as a function of normalized position $x$, measured upward from the loop base, with $T_o/T_{\rm max}$ set to the value 0.01 in the case of turbulence-dominated conduction. For turbulence-dominated conduction, the weaker (in fact, inverse) dependence of the conduction coefficient $\kappa_o$ on temperature leads to steeper (shallower) temperature gradients at high (low) temperatures, compared to the case of collision-dominated conduction. Physically, this can be explained as follows: in the face of reduced conductivity in the hot upper regions of the loop, the heat flux can only transport excess energy (heating - radiation) from the corona by steepening the temperature gradient. This reduced flux of energy entering the lower (and now more conductive than in the collisional case) atmosphere can be supported by a much shallower temperature gradient in the lower regions of the loop.

In contrast to the results of \cite{2010ApJ...714.1290M}, which showed (his Figure~3) that the loop temperature profiles were largely independent of the value of the parameter $\alpha$ (the temperature dependence of the loop heating rate -- Equation~(\ref{dimensionless-parameters})) -- the results of Figure~\ref{t-profiles} show that the loop temperature profile is {\it very} sensitive to the value of $\delta$; i.e., to the nature of the thermal conductive transport term.

\subsection{Hybrid Collisional-Turbulent Loop Model}\label{hybrid_model}

For low values of the temperature (i.e., near the base of the loop), the ratio of collisional to turbulent mean free paths $R \equiv \lambda_C/\lambda_T \ll 1$, and hence the usual assumption of collision-dominated thermal conduction is valid. However, at the high temperatures and low densities found near the loop apex, the collisional mean free path $\lambda_C$ is much larger\footnote{It is interesting to note that the expansion parameter $\lambda/L$, commonly called the Knudsen number, must, by definition, be small for a fluid (rather than particle-based) treatment of thermal conductive transport to be valid. Collisional path lengths of this magnitude for thermal electrons thus imply that common formulations of thermal conduction in the fluid limit have often been applied beyond their limits of applicability.} (e.g., for $T = 3 \times 10^6$~K and $n = 10^{9}$~cm$^{-3}$, $\lambda_C \simeq 2 \times 10^8$~cm), so that for turbulent mean free paths $\lambda_T$ smaller than this, conductive transport is driven predominantly  by turbulent scattering. A physically-correct loop model thus requires the merging of a collision-dominated conduction profile at low temperatures with a turbulence-dominated conduction profile at higher temperatures.

The base temperature $T_o$ (and hence value of $\zeta_o$; Equations~(\ref{solution-upper}) and~(\ref{solution-upper-boundary})) associated with turbulent conduction solution can now be given a physical significance: it is the temperature at which the collisional and turbulent mean free paths are comparable; i.e., $\lambda_T=\lambda_C \, (T=T_o)$.  Since the collisional mean free path scales as $T^2/n \sim T^3/P$, for a uniform pressure loop

\begin{equation}\label{lambdaCTo}
\lambda_C(T_o) = \lambda_C(T=T_{\rm max}) \left ( \frac{T_o}{T_{\rm max}} \right )^3 \,\,\, .
\end{equation}
But, by definition of the interface temperature $T_o$, $\lambda_C(T_o) = \lambda_T$; thus

\begin{equation}\label{eta-o-def}
\frac{\lambda_T}{\lambda_C(T=T_{\rm max})} = \left ( \frac{T_o}{T_{\rm max}} \right )^3 \,\,\, ,
\end{equation}
so that the value of $\zeta_o = (T/T_{\rm max})^{1/2}$ corresponding to the base of the turbulent region is

\begin{equation}\label{value-of-zetao}
\zeta_o =  \left ( \frac{\lambda_T}{\lambda_C(T=T_{\rm max})} \right )^{1/6} \,\,\, .
\end{equation}
Similarly, the collision-dominated solution~(\ref{solution-case-1}) is valid from $\eta = 0$ up to a value

\begin{equation}\label{value-of-etao}
\eta_o \equiv \zeta_o^7 = \left ( \frac{\lambda_T}{\lambda_C(T=T_{\rm max})} \right )^{7/6} \,\,\, .
\end{equation}
Using Equation~(\ref{mean-free-path-collisional}) for the collisional mean free path $\lambda_C$ gives

\begin{equation}\label{etao-determination}
\eta_o = \left ( \frac{2 \pi e^4 \Lambda n \, \lambda_T}{4 k_B^2 T_{\rm max}^2} \right )^{7/6} = \left ( \frac{\pi e^4 \Lambda \, \lambda_T \, P}{4 k_B^3 T_{\rm max}^3} \right )^{7/6} = \left ( \frac{\kappa_{oT}}{\kappa_{oc} \, T_{max}^3} \right)^{7/6} = \left ( \frac{\lambda_T \, P}{c_R \, T_{\rm max}^3} \right )^{7/6} \,\,\, ,
\end{equation}
(where we have used Equation~(\ref{kappa-ratio})), and hence to an interface temperature

\begin{equation}\label{To-determination}
T_o = \left ( \frac{\lambda_T \, P}{c_R} \right )^{1/3} \,\,\, .
\end{equation}
The value of the interface temperature $T_o$, that marks the upper/lower levels of the collisional/turbulent domains, respectively, is thus set by the values of the loop pressure $P$ and the turbulent scattering length $\lambda_T$.

In summary, the temperature profile for a collisional-turbulent hybrid physical model is given by Equations~(\ref{first-integral}) and~(\ref{first-integral-turbulent}), viz.

\begin{eqnarray}\label{summary-split-solution}
\frac{\epsilon}{2} \, {\eta^{\prime}}^2 &=& \frac{7}{2} \, \eta^{2/7} - \xi \,  \eta \qquad \qquad \qquad \qquad ; \, \eta \le \eta_o
\cr
\frac{\epsilon_T}{2} \, {\zeta^{\prime}}^2 &=& \frac{1}{4} \left ( 1 - \frac{1}{\zeta^4} \right ) + \xi \, (1 -  \zeta ) \qquad \quad ; \, \zeta \ge \zeta_o = \eta_o^{1/7} \,\,\, ,
\end{eqnarray}
where $\eta = (T/T_{\rm max})^{7/2}$, $\zeta = (T/T_{\rm max})^{1/2}$, and the interface value $\eta_o$ is given by Equation~(\ref{etao-determination}).

The value of the dimensionless parameter $\xi$ can no longer be obtained by appealing to the condition $\zeta^\prime=0$ at $\zeta=1$, similar to the condition used in obtaining the result~(\ref{xi-collisional}) for a wholly collisional solution. Instead, its value is obtained by matching the conductive flux at the interface between the collisional solution in the lower part of the loop and the turbulent solution in the upper part of the loop. At this interface, the temperature variable $\eta = \eta_o$, and the collisional and turbulent fluxes are given by

\begin{equation}\label{flux-match-at-boundary}
F_{\rm coll} = \frac{2}{7} \, \kappa_{oc} \, \frac{T_{\rm max}^{7/2}}{L} \, \eta_o^\prime \, ; \qquad F_{\rm turb} = 2 \, \kappa_{oT} \, \frac{T_{\rm max}^{1/2}}{L} \, \zeta_o^\prime \,\,\, ,
\end{equation}
respectively. Equating the (square of) these fluxes at $\eta=\eta_o$ ($\zeta=\zeta_o \equiv \eta_o^{1/7}$) and using the temperature profile solutions~(\ref{summary-split-solution}) for each part of the loop gives

\begin{equation}\label{derivative-match}
\frac{\epsilon_T}{\epsilon} \, \left ( \frac{7}{2}  \eta_o^{2/7} - \xi \, \eta_o \right ) = 49 \, \left ( \frac{\kappa_{oT}}{\kappa_{oc} T_{\rm max}^3} \right )^2 \left [ \frac{1}{4} \left ( 1 - \frac{1}{\eta_o^{4/7}} \right ) + \xi \left (1 - \eta_o^{1/7} \right ) \right ] \,\,\, .
\end{equation}
But, from the definitions of $\epsilon$ and $\epsilon_T$ (Equations~(\ref{dimensionless-parameters}) and~(\ref{dimensionless-parameters-turbulent})),

\begin{equation}\label{epsilon-ratio}
\frac{\epsilon_T}{\epsilon} = 7 \left ( \frac{\kappa_{oT}}{\kappa_{oc} T_{\rm max}^3} \right ) = 7 \, \eta_o^{6/7} \,\,\, ,
\end{equation}
where we have used Equation~(\ref{etao-determination}). Hence

\begin{equation}\label{derivative-match-dimensionless}
\frac{7}{2}  \eta_o^{2/7} - \xi \, \eta_o = 7 \, \eta_o^{6/7} \, \left [ \frac{1}{4} \left ( 1 - \frac{1}{\eta_o^{4/7}} \right ) + \xi \left (1 - \eta_o^{1/7} \right ) \right ]\,\,\, .
\end{equation}

Grouping the terms involving the parameter $\xi$ results in an explicit expression for that parameter:

\begin{equation}\label{value-of-xi-physical-model}
\xi = \frac{7}{2} \, \frac{ \eta_o^{2/7} - \frac{1}{2} \, \eta_o^{6/7} \, \left (1 - \eta_o^{-4/7}  \right ) }{\eta_o + 7 \, \eta_o^{6/7} \, \left (1 - \eta_o^{1/7}  \right )} = \frac{7}{4 \, \eta_o^{4/7}} \, \left ( \frac{ 3 - \eta_o^{4/7}}{7 - 6 \eta_o^{1/7} } \right ) \,\,\, .
\end{equation}
As the collisional/turbulent interface level approaches the loop apex ($\eta_o \rightarrow 1$), $\xi \rightarrow 7/2$, consistent with the value obtained earlier (Equation~(\ref{xi-collisional})) for the wholly collisional solution. On the other hand, when turbulent conduction dominates throughout most of the loop (corresponding to small values of $\eta_o$), $\xi \rightarrow 3/{4 \eta_o^{4/7}}$. The value of $\xi$ therefore increases as $\eta_o$ decreases, i.e., as more and more of the loop is dominated by turbulent scattering.

Such an increased value of $\xi$ corresponds (Equation~(\ref{dimensionless-parameters})) to a greater amount of heat input $H$. As we shall see in Section~\ref{dem-profiles-section}, such an increased amount of heat input is necessary to sustain the greater amount of radiative losses from the loop, especially from the lower-temperature regions of the loop, which have an increased amount of emitting material in a given temperature range, due to the shallower temperature gradients in that region (cf. Figure~\ref{t-profiles}).

The vanishing of the left-hand side of each of the energy equations~(\ref{eta-double-prime}) and~(\ref{eta-double-prime-turbulent}) (corresponding to the wholly collisional and wholly turbulent cases, respectively) corresponds to the point where heat input and radiative losses balance locally, and hence the divergence of the conductive flux changes sign, i.e., when thermal conduction transitions from a cooling mechanism to a heating mechanism. Using Equations~(\ref{dimensionless-vars}) and~(\ref{dimensionless-parameters-turbulent}), we find that the left-hand sides of both energy equations equal zero when

\begin{equation}\label{cooling-heating-transition}
\theta \equiv \frac{T_{\rm transition}}{T_{\rm max}} = \xi^{-2/5} \,\,\, ;
\end{equation}
this result follows straightforwardly from the general energy equation~(\ref{energy-equation-general}) and the definition of $\xi$ (Equation~(\ref{dimensionless-parameters})). For the wholly collisional case, $\xi = 7/2$ (Equation~(\ref{xi-collisional})) and so $\theta \simeq 0.6$; for temperatures above this ``transition temperature'' conduction is a cooling term balancing the excess heating over radiation, while for temperatures below the transition temperature conduction is a heating term supplying additional energy to be radiated away. As the role of turbulence in the upper regions of the loop becomes more pronounced, the value of the heating parameter $\xi$ increases and Equation~(\ref{cooling-heating-transition}) then shows that the value of $\theta$ decreases below its collisional value $\simeq 0.6$.

In summary, the more predominant role of turbulence associated with low values of the turbulent mean free path $\lambda_T$ leads to:

\begin{itemize}

\item lower values of the collisional/turbulent interface temperature (Equation~(\ref{To-determination})); and

\item higher values of the heating parameter $\xi$ (Equation~(\ref{value-of-xi-physical-model})).

\end{itemize}

Consistent with this higher level of overall heating, thermal conduction acts as a cooling mechanism throughout a greater fraction of the loop, which in turn leads to

\begin{itemize}

\item lower values of the heating/cooling transition temperature $\theta$ (Equation~(\ref{cooling-heating-transition})).

\end{itemize}

\begin{figure}[pht]
	\centering
	\includegraphics[width=0.6\linewidth]{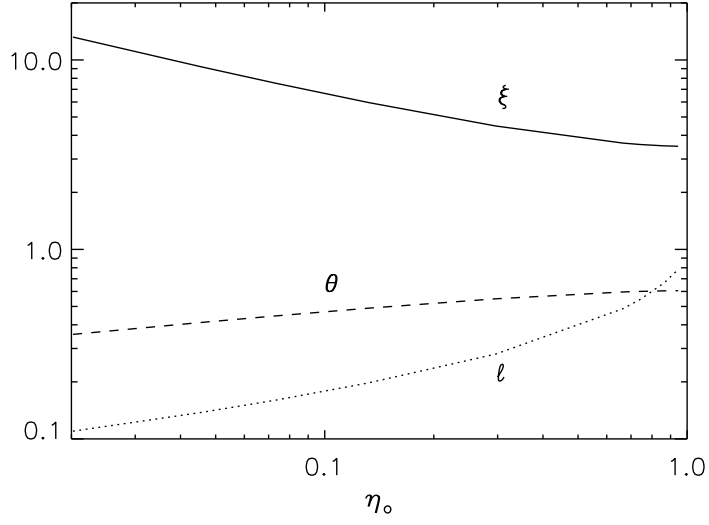}
	\caption{\label{multi-plot} Variation of the heating parameter $\xi$ (Equation~(\ref{value-of-xi-physical-model})), the collisional/turbulent interface position $\ell$ (Equations~(\ref{solution-upper}) and~(\ref{solution-both-halves-at-boundary})), and the cooling/heating boundary temperature $\theta$ (Equation~(\ref{cooling-heating-transition})), all as functions of the quantity $\eta_o$ (Equation~(\ref{etao-determination})), that parameterizes the temperature of the collisional/turbulent interface.  Asymptotic behaviors of all three parameters as $\eta_o \rightarrow 0$ and $\eta_o \rightarrow 1$ are discussed in the text.}
\end{figure}

These trends are depicted pictorially in Figure~\ref{multi-plot}, which shows the behavior of the heating parameter $\xi$ (Equation~(\ref{value-of-xi-physical-model})), the collisional/turbulent interface position $\ell$ (measured upward from the loop footpoint and normalized to the loop half-length $L$; Equations~(\ref{solution-upper}) and~(\ref{solution-both-halves-at-boundary})), and the cooling/heating boundary temperature $\theta$ (Equation~(\ref{cooling-heating-transition})), all as functions of the quantity $\eta_o$ (Equation~(\ref{etao-determination})) that parameterizes the temperature of the collisional/turbulent interface. As $\eta_o \rightarrow 1$, more and more of the loop is governed by collision-dominated conduction, and $\xi \rightarrow 7/2$ (Equation~(\ref{xi-collisional})), $\ell \rightarrow 1$, and the transition temperature variable $\theta \rightarrow (7/2)^{-2/5} \simeq 0.6$.  At the other extreme, as $\eta_o \rightarrow 0$, more and more of the loop becomes governed by conductive heat transport driven by turbulent scattering; the heating parameter grows like $\eta_o^{-4/7}$ (Equation~(\ref{value-of-xi-physical-model})), the value of $\ell$ approaches zero, and the transition temperature variable $\theta$ slowly approaches zero like $\xi^{-2/5} = \eta_o^{8/35}$.

Using Equations~(\ref{dimensionless-vars}) and~(\ref{dimensionless-parameters}) for $\eta_o$ and $\xi$, respectively, in the relation $\xi = 3/{4 \eta_o^{4/7}}$ gives

\begin{equation}\label{scaling-law-2-xi}
\frac{H \, (2 k_B)^{2} \, T_{\rm max}^{5/2}}{\chi_o \, P^{2}} = \frac{3}{4} \left ( \frac{T_o}{T_{\rm max}} \right )^{-2} \,\,\, ,
\end{equation}
or

\begin{equation}\label{H-Tmax-turb}
H = \frac{3 \, \chi_o P^2}{16 (k_B T_o)^2 \, T_{\rm max}^{1/2}} \,  \,\,\, .
\end{equation}
Comparison with the equivalent result~(\ref{scaling-law-2-Martens-final}) for the collisional case shows how the loop base temperature $T_o$ now explicitly plays a role in determining the magnitude of the volumetric heating function $H$ \citep[cf.][]{2019ApJ...880...80B}. The relation $\xi = 3/{4 \eta_o^{4/7}}$ also implies that as $\eta_o \rightarrow 0$, $\xi \rightarrow \infty$, so that the denominator in the integrand in Equation~(\ref{solution-upper-boundary}) is dominated by the second term under the radical. Substituting $\zeta_o = 0$ and $\ell = 0$ in this limit then gives

\begin{equation}\label{matching-solutions-right-side-small-etao}
\sqrt{\frac{2}{\epsilon_T}} = \frac{1}{\sqrt{\xi}} \, \int_0^1 \frac{d \zeta}{\sqrt{1 - \zeta}} = \frac{2}{\sqrt{\xi}} \,\,\, ,
\end{equation}
from which

\begin{equation}\label{matching-solutions-right-side-small-etao-squared}
\epsilon_T = \frac{\xi}{2} = \frac{3}{8 \, \eta_o^{4/7}} \,\,\, .
\end{equation}
Using Equations~(\ref{dimensionless-vars}) and~(\ref{dimensionless-parameters-turbulent}), this may be written as

\begin{equation}\label{epsilon-T-whole-loop-2}
\frac{2 \, \kappa_{oT} \, (2 k_B)^2 \, T_{\rm max}^{3}}{\chi_o \, P^2 \, L^2} = \frac{3}{8} \, \left ( \frac{T_{\rm max}}{T_o} \right )^{2} \,\,\, ,
\end{equation}
and substituting for $\kappa_{oT}$ from Equation~(\ref{kappa-ratio}) gives the ``first scaling law''

\begin{equation}\label{scaling-law-1}
T_{\rm max} = \frac{3 \, \chi_o c_R}{64 \, \kappa_{oc} \, \lambda_T \, (k_B T_o)^2} \, P \, L^2 \simeq \frac{7.3 \times 10^5}{\lambda_T \, T_o^2} \, P \, L^2 \,\,\, .
\end{equation}
This result improves (by a factor of 3) the approximate scaling law of \cite{2019ApJ...880...80B} (their Equation~(23)). Then using Equation~(\ref{scaling-law-1}) for $T_{\rm max}$ in Equation~(\ref{H-Tmax-turb}) gives the ``second scaling law''

\begin{equation}\label{scaling-law-2}
H = \left ( \frac{3 \, \kappa_{oc} \, \chi_o \lambda_T}{4 c_R} \right )^{1/2} \, \frac{1}{k_B T_o} \, P^{3/2} \, L^{-1} \simeq \frac{1.9 \times 10^9 \, \lambda_T^{1/2}}{T_o} \, \frac{P^{3/2}}{L} \,\,\, ,
\end{equation}
which improves (by a factor $\sqrt{3}$) the approximate scaling law derived in \cite{2019ApJ...880...80B} (their Equation~(24)), and again highlights the role played by the base temperature $T_o$ in driving the volumetric heating rate $H$. In a future work, we intend to critically compare the scaling law~(\ref{scaling-law-1}), and its collisional counterpart~(\ref{scaling-law-1-Martens-final}), with observations, and we encourage others to do the same. Such comparisons will allow useful constraints to be made on the value of the turbulence scale length $\lambda_T$ (given reasonable estimates of the base temperature $T_o$), and then in turn, through Equation~(\ref{scaling-law-2}), on the heating rate $H$.

We now have all the information needed to construct the temperature profile for the hybrid collisional/turbulent model. For prescribed values of the quantities $\lambda_T$, $P$, and $T_{\rm max}$, the value of the interface temperature parameter $\eta_o$ follows from Equation~(\ref{etao-determination}), recalling that the constant $c_R \simeq 3.15 \times 10^{-12}$~erg~cm$^{-2}$~K$^{-3}$  \cite[Equation~(11) of][]{2019ApJ...880...80B}. Substituting this value of $\eta_o$ in Equation~(\ref{value-of-xi-physical-model}) then gives the corresponding value of the heating-rate-related quantity $\xi$ (Equation~(\ref{dimensionless-parameters})). Equations~(\ref{summary-split-solution}) can now be directly integrated to give the temperature parameter $\eta$ (Equation~(\ref{dimensionless-vars})) as an (implicit) function of the dimensionless position $x$:

\begin{eqnarray}\label{solution-both-halves}
\int_0^{\eta} \left ( \eta^{2/7} -  \xi \, \eta \right )^{-1/2} \, d \eta &=& \sqrt{\frac{7}{\epsilon}} \, x \, ; \qquad 0 < x < \ell \cr
\int_{\eta_o^{1/7}}^{\eta^{1/7}} \left ( \frac{1}{4} \, \left( 1 - \frac{1}{\zeta^4} \right) + \, \xi \,  ( 1 - \zeta ) \right )^{-1/2} \, d \zeta & = & \sqrt{\frac{2}{\epsilon_T}} \,\, (x - \ell) ; \qquad \ell < x < 1 \,\,\, .
\end{eqnarray}
We can now apply these at the upper boundary of the collision-dominated domain ($\eta=\eta_o$; $x = \ell$) and at the loop apex ($\eta =1$; $x = 1$), respectively, to obtain

\begin{eqnarray}\label{solution-both-halves-at-boundary}
\int_0^{\eta_o} \left ( \eta^{2/7} - \xi \, \eta \right )^{-1/2} \, d \eta &=& \sqrt{\frac{7}{\epsilon}} \, \ell \, \cr
\int_{\eta_o^{1/7}}^{1} \left ( \frac{1}{4} \, \left( 1 - \frac{1}{\zeta^4} \right) + \, \xi \,  ( 1 - \zeta ) \right )^{-1/2} \, d \zeta & = & \sqrt{\frac{2}{\epsilon_T}} \,\, (1 - \ell)
\end{eqnarray}
for the lower (collision-dominated) and upper (turbulence-dominated) regions of the loop, respectively. Dividing these two results gives

\begin{equation}\label{solution-match-result}
\frac{\int_0^{\eta_o} \left ( \eta^{2/7} - \xi \, \eta \right )^{-1/2} \, d \eta }{\int_{\eta_o^{1/7}}^{1} \left ( \frac{1}{4} \, \left( 1 - \frac{1}{\zeta^4} \right) + \, \xi \,  ( 1 - \zeta ) \right )^{-1/2} \, d \zeta}  = \sqrt{\frac{7}{2}} \, \sqrt{\frac{\epsilon_T}{\epsilon}} \, \frac{\ell}{1-\ell} = \frac{7}{\sqrt{2}} \, \eta_o^{3/7} \, \frac{\ell}{1-\ell} \,\,\, ,
\end{equation}
where in the last equality we have used Equation~(\ref{epsilon-ratio}). This result provides the value  of the (dimensionless) boundary position $\ell$ as a function of the interface temperature variable $\eta_o$, determined straightforwardly from the physical parameters of the loop using Equation~(\ref{etao-determination}). The value of the scaling parameters $\epsilon$ and $\epsilon_T$ (Equations~(\ref{dimensionless-parameters}) and~(\ref{dimensionless-parameters-turbulent}), respectively) then follow from the first and second of Equations~(\ref{solution-both-halves-at-boundary}), respectively, and, finally, the temperature profile $\eta(x)$ then follows from solving Equations~(\ref{solution-both-halves}) in each segment.

\begin{figure}[pht]
\centering
\includegraphics[width=0.6\linewidth]{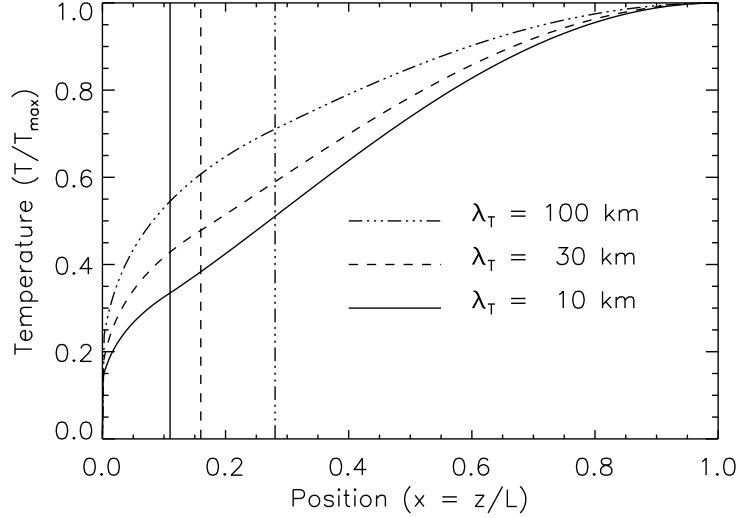}
\caption{\label{t-profile-physical-model} Normalized temperature profiles for the hybrid physical model, with $T_{\rm max} = 3 \times 10^6$~K, $P = 3.0$~dyne~cm$^{-2}$, and the values of $\lambda_T$ indicated. The vertical dashed lines indicate the normalized positions $\ell$ of the interface between the collisional and turbulent regimes; for $x < \ell$ conduction is collision-dominated, while for $x > \ell$ it is turbulence-dominated.}
\end{figure}

Figure~\ref{t-profile-physical-model} shows the resulting temperature profiles, for various values of the turbulent mean free path $\lambda_T$, and for a loop with apex temperature $T_{\rm max} = 3 \times 10^6$~K and pressure $P = 3.0$~dyne~cm$^{-2}$. The values of the (normalized) position $\ell$ at which the conduction transitions from collision-dominated (small $x$) to turbulence-dominated (large $x$) are shown as vertical lines in the Figure. As $\lambda_T$ decreases, so does the value of $\ell$, corresponding to the turbulence-dominated region extending over an ever-increasing part of the loop.  And, as we noted in the results for a loop in which turbulent scattering dominates throughout (Figure~\ref{t-profiles}), the general tendency is for loops with a lower value of $\lambda_T$ to have steeper temperature gradients at high temperatures, and shallower temperature gradients at low temperatures.

\section{Differential Emission Measure Profiles}\label{dem-profiles-section}

The differential emission measure (cm$^{-5}$~K$^{-1}$) is defined as

\begin{equation}\label{DEM-define}
DEM(T) = n^2 \frac{dz}{dT} = \frac{P^2}{(2 k_B)^2 \, T^2} \, \frac{dz}{dT} \,\,\, .
\end{equation}
It measures the ``radiating potential'' in a given temperature range, such that convolution of the $DEM(T)$ profile with the emissivity function $G(\lambda;T)$ expressing the emission spectrum per unit wavelength $\lambda$ at temperature $T$ gives the total emitted spectrum per unit wavelength:

\begin{equation}
I(\lambda) = \int_{T=0}^\infty DEM(T) \, G(\lambda; T) \, dT \,\,\, .
\end{equation}
Using Equation~(\ref{dimensionless-vars}), Equation~(\ref{DEM-define}) can be written as

\begin{equation}\label{DEM-eta}
DEM(T[\eta]) = \frac{P^2 \, L}{4 \, k_B^2 \, T_{\rm max}^3} \, \left [ \frac{7}{2} \, \frac{\eta^{1/7}}{\eta^\prime} \right ] \,\,\, ,
\end{equation}
where $\eta^\prime$ denotes the derivative of $\eta$ with respect to the dimensionless distance $x=z/L$, and so is readily evaluated from the $\eta(x)$ profiles obtained above. At the coolest temperatures in the loop, the collision-dominated Equation~(\ref{first-integral-case-1}) applies, and can further be well approximated as

\begin{equation}\label{first-integral-low-eta}
\eta^{\prime} = \sqrt{\frac{7}{\epsilon}} \, \eta^{1/7}  \,\,\, ,
\end{equation}
so that the $DEM$ tends to a constant value

\begin{equation}\label{DEM-low-temp-limit}
DEM = \sqrt{\frac{7 \, \epsilon}{4}} \, \frac{P^2 \, L}{4 \, k_B^2 \, T_{\rm max}^3} = \sqrt{\frac{\kappa_{oc}}{8 \chi_o \, k_B^2}} \, P \simeq 8 \times 10^{21} \, P \,\,\, ,
\end{equation}
where we have used Equation~(\ref{dimensionless-parameters}). This relation also, obviously, applies\footnote{Although \cite{2010ApJ...714.1290M} did not provide $DEM$ profiles corresponding to his loop temperature profiles, those temperature profiles (his Figure~3) were so similar that we would expect the $DEM$ profiles to be very insensitive to the value of the parameter $\alpha$, the temperature dependence of the loop heating rate.  Thus observations of spectral line intensities would not provide a clear indicator of the value of $\alpha$.} for loops in which conduction is collision-dominated throughout \citep[e.g.,][]{2010ApJ...714.1290M}. To the best of our knowledge, this predicted (constant) behavior of $DEM$ at low temperatures in active region loops for which classical \citep{1962pfig.book.....S} conduction applies has not been noted previously in the literature.

\begin{figure}[pht]
\centering
\includegraphics[width=0.6\linewidth]{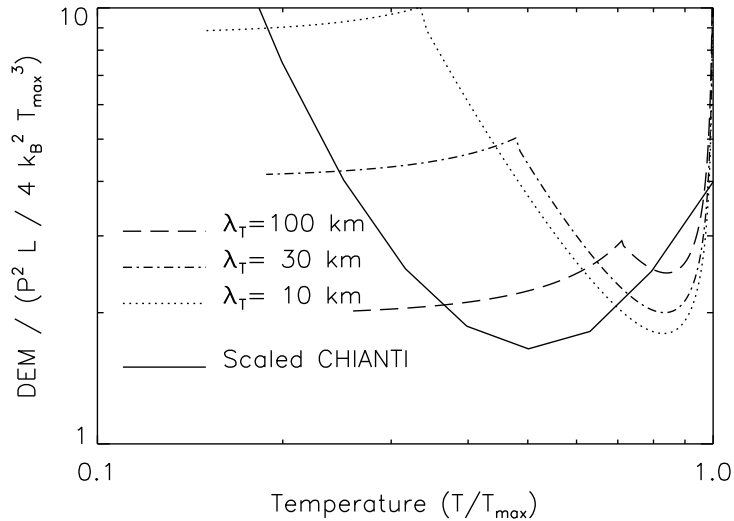}
\caption{Normalized differential emission measure profiles for a physical model with collision-dominated conduction in the lower part of the loop and turbulence-dominated conduction in the upper portion, with $T_{\rm max} = 3 \times 10^6$~K, $P = 3.0$~dyne~cm$^{-2}$, and the values of $\lambda_T$ indicated. The corresponding scaled CHIANTI $DEM$ profile inferred from observations is shown for comparison (see text for details).}\label{dem-profiles}
\end{figure}

Figure~\ref{dem-profiles} shows the $DEM$ structures for several values of the turbulence mean free path $\lambda_T$, scaled by the constant term in Equation~(\ref{DEM-eta}). For classical conduction, the high power of $T$ in the thermal conduction coefficient $\kappa_{oc}$ forces very steep temperature gradients (and so relatively small amounts of material per unit temperature) at low temperatures (Figure~\ref{t-profiles}).  However, for models that include significant turbulent conduction (whether throughout the entire loop or only in the upper regions), the temperature profile is much flatter at low temperatures (Figures~\ref{t-profiles} and~\ref{t-profile-physical-model}, respectively). This, combined with the increased density at such low temperatures ($n \propto P/T$), gives a significant rise in $DEM$ as the temperature decreases. It should be noted that even though the lower portion of the loop is collision-dominated, the requirement that it continuously link to the turbulence-dominated upper part of the loop changes the $DEM$ structure from that corresponding to a loop in which collisional conduction dominates throughout.

\begin{figure}[pht]
	\centering
	\includegraphics[width=0.6\linewidth]{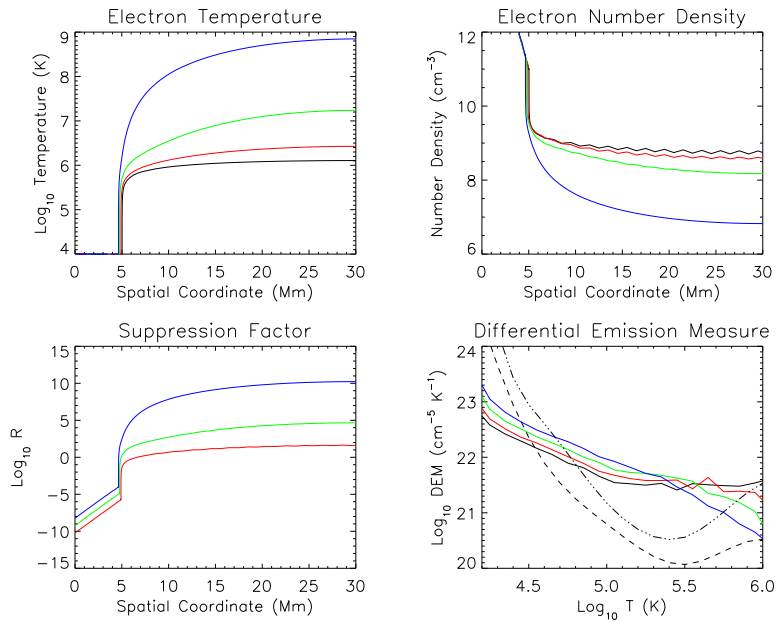}
	\caption{Hydrostatic equilibrium solutions for turbulence with spatially uniform scales $\lambda_T = 100$~km (red), 10~km (green), and 1~km (blue). Each solution was found by relaxation from an initially collisional state ($\lambda_T = \infty$; black). The dashed and dot-dashed lines shows the empirical CHIANTI differential emission measure curves for the quiet Sun and for active regions; see text for discussion.}\label{numerical-dem-profiles}
\end{figure}

Such an enhancement in the differential emission measure at low temperatures compared to the values predicted using classical heat conduction is indeed ubiquitously observed \cite[e.g.,][]{1981ApJ...247..686R}. \cite{2002ApJS..139..281L} have used data from the Solar Ultraviolet Measurements of Emitted Radiation (SUMER) spectrometer \citep{1995SoPh..162..189W} on board the Solar and Heliospheric Observatory (SoHO) to deduce empirical $DEM$ curves for different spectral lines; sample $DEM$ results are shown in their Figures~2 through~5. These empirical $DEM$ curves generally have a concave upward structure with a pronounced minimum, similar to the model $DEM$ results in Figures~\ref{dem-profiles} and~\ref{numerical-dem-profiles} (and, we note parenthetically, completely inconsistent with the low-temperature $DEM$ behavior of Equation~(\ref{DEM-low-temp-limit}) for a model loop in which conduction is dominated by collisions throughout). \cite{1986ApJ...301..440A} discuss the failure of static\footnote{In this context, ``static'' models must also be interpreted as those which include time-dependent heating that repeats on short timescales \citep[high-frequency heating - see, e.g.,][]{2011ApJ...742L...6M}.} (and steady-state flow) coronal loop models to reproduce this observed sharp rise in the emission measure at temperatures from 100,000~K down to 20,000~K, despite the models and observations being in good agreement above 100,000~K. They note that although several solutions to this inconsistency have been proposed, these all rely on different mechanisms operating above and below 100,000~K, with no compelling physical reason why this should be so, or why 100,000~K is the critical temperature. Another possibility noted is that some key physics (such as kinetic effects) is missing from static/steady models, but \cite{1986ApJ...301..440A} state that ``$\cdots$ it has yet to be demonstrated that the kinetic effects are sufficient to account for all of the large discrepancy between the observed and the predicted line fluxes.''  A possible explanation for the excess of cool material is that individual loop strands undergo heating and cooling cycles, with the heating and cooling timescales being such that a significant amount of cool material is always present. However, the results in Section~\ref{dem-profiles-section} show that a relatively straightforward physical mechanism, namely the suppression of thermal conduction by turbulent scattering at sub-collisional (kinetic) scales, results naturally in changes to the overall loop temperature profile, and hence the differential emission measure at low temperatures, that are much more consistent with observations.

An overall $DEM$ profile for an active region loop has been computed by K. Dere, based on the average active region model atmosphere of \cite{1978ApJS...37..485V}, with an assumed uniform pressure of $nT = 3 \times 10^{15}$~cm$^{-3}$~K (i.e., $P \simeq 0.8$~dyne~cm$^{-2}$) and a maximum temperature $T_{\rm max} \simeq 10^{6.2}$~K. To compare this $DEM$ profile with those of the models (Figure~\ref{dem-profiles}) requires that we calculate the scaling quantity $P^2 L/4 k_B^2 T_{\rm max}^3$, and this requires a value for the loop half-length $L$. Using the \cite{1978ApJ...220..643R} scaling law~(\ref{scaling-law-1-Martens-final}) with $P = 0.8$~dyne~cm$^{-2}$ and $T_{\rm max} = 10^{6.2}$~K yields a fairly large loop half-length $L \simeq 2 \times 10^9$~cm. However, given that the scaling law~(\ref{scaling-law-1-Martens-final}) is not valid for models that incorporate turbulent conduction, we used a lower value $L = 7 \times 10^8$~cm (10~arcseconds) and a value $T_{\rm max} = 3 \times 10^6$~K (to match\footnote{Note that with $P=0.8$~dyne~cm$^{-2}$ and $L = 7 \times 10^8$~cm, the \cite{1978ApJ...220..643R} scaling law~(\ref{scaling-law-1-Martens-final}) gives a rather low (quiet-Sun-level) peak temperature $T_{\rm max} \simeq 1.2 \times 10^6$~K. For comparison, however, the fully-turbulent scaling law~(\ref{scaling-law-1}), with $P = 0.8$~dyne~cm$^{-2}$, $L = 7 \times 10^8$~cm, base temperature $T_o = 10^5$~K, and a turbulent scale length $\lambda_T = 10^7$~cm (100~km) gives $T_{\rm max} \simeq 3 \times 10^6$~K, consistent with the value used. Lower values of $\lambda_T$ and/or $T_o$ yield (Equation~(\ref{scaling-law-1})) even higher values for $T_{\rm max}$, so that the choice $T_{\rm max} = 3 \times 10^6$~K would now fall between the peak temperatures appropriate to the fully-collisional and fully-turbulent regimes, respectively.} the model calculations), resulting in a $DEM$ scaling quantity $P^2 L/4 k_B^2 T_{\rm max}^3 = 2 \times 10^{20}$. The CHIANTI $DEM$ profile, scaled (downward) by this factor, is superimposed on the model results of Figure~\ref{dem-profiles}. We see that the value of the minimum in the $DEM$ profile is close to that of the models; however, the models predict a more localized minimum at a somewhat larger temperature.

Both as a ``reality check'' on the above modeling, and to form a basis for comparison, we also calculated the temperature structure of loops by numerically solving the energy equation~(\ref{energy-equation-general}) with a continuously varying conduction term representative of the local conditions at each point. Figure~\ref{numerical-dem-profiles} shows results obtained using the HYDRAD code \citep{2013ApJ...770...12B}, which solves the multi-fluid hydrodynamic equations in the field-aligned direction, allowing for changes in the flux tube cross-section and for gravitational acceleration, and including key plasma physics processes such as thermal conduction, viscous interactions, inter-species collisions, and radiation. The relative importance of collisions and turbulence to thermal conduction is allowed to change continuously, depending upon the local collisionality of the plasma, via the relations

\begin{equation}\label{lambda-general}
\nu = \nu_C + \nu_T \, ; \qquad \frac{1}{\lambda} = \frac{1}{\lambda_C} + \frac{1}{\lambda_T}
\end{equation}
for the scattering frequencies $\nu$ and corresponding mean free paths $\lambda$, with the conduction coefficient $\kappa$ proportional to $\lambda$ (cf. Equations~(\ref{conductive-parameters-coll}) and~(\ref{conductive-parameters-turbulent})). The field-aligned electron temperature and number density are shown in the top two panels for a range of (uniform) values of $\lambda_T$. The upper-left plot shows that for turbulent mean free paths less than 1~km, unreasonably high coronal temperatures of order $10^9$~K; this effectively imposes a lower limit on the value of $\lambda_T$. The lower-left plot shows values of the ratio $R=\lambda_C/\lambda_T$ (cf. Equations~(\ref{mean-free-path-collisional}) and~(\ref{conductive-parameters-turb})) throughout the loop, showing that thermal conduction can be hugely suppressed (by factors as large as $10^{10}$) in the hot, tenuous corona, whereas the footpoint regions are still strongly collision-dominated. The lower-right plot shows the $DEM$ profiles obtained, with the CHIANTI active-region and quiet-Sun $DEM$ profiles superimposed. The HYDRAD models produce a $DEM$ profile that is characterized by a rather broad minimum compared to the empirical CHIANTI profile, with significant (up to $\sim$1.5 orders of magnitude) excess emission in the range $4.8 \lapprox \log T \lapprox 5.8$. This is in contrast to the analytic models of Figure~\ref{dem-profiles}, which are characterized by a much {\it narrower} minimum than the empirical CHIANTI profile. We emphasize, however, that the general excess of emission at relatively cool temperatures is a feature common to all three models: the analytic models of Figure~\ref{dem-profiles}, the numerical models of Figure~\ref{numerical-dem-profiles}, and the empirical CHIANTI profile, and that this is a feature that the collision-dominated conduction result (Equation~(\ref{DEM-low-temp-limit})) completely fails to reproduce.

\section{summary and conclusions}\label{conclusions}

We have seen that the (likely) presence of turbulence in active region solar loops fundamentally changes the characteristics of heat transport by thermal conduction, affecting not only the loop scaling laws \cite{2019ApJ...880...80B} but also the temperature structure throughout the loop. Compared to temperature profiles \citep{2010ApJ...714.1290M} based on classical (i.e., collision-driven) conduction \citep{1962pfig.book.....S}, temperature profiles in loops where conductive heat transport is dominated by turbulent scattering have steeper temperature  gradients in the high corona and shallower temperature gradients in the low corona (Figure~\ref{t-profiles}). Physically, this is a result of the much weaker dependence of the heat conduction coefficient $\kappa$ on temperature (Equations~(\ref{conductive-parameters-collisional}) and~(\ref{conductive-parameters-turbulent})). Since the differential emission measure is inversely proportional to the temperature gradient, the $DEM$ profiles in loops where turbulent scattering dominates the thermal heat transport, show significant enhancements at low temperatures compared to loops where collisional scattering dominates the thermal heat transport (Figure~\ref{dem-profiles}).

In practice the temperature and density variation along a real loop will likely result in a situation where turbulence-driven conduction dominates at high temperatures and collision-dominated conduction dominates at low temperatures. Consideration of such a ``hybrid'' loop model shows, both through matching of collisional and turbulent solutions (Figure~\ref{t-profile-physical-model}), and through more precise numerical modeling involving a continuously variable ratio of the collisional to turbulent mean free paths (Figure~\ref{numerical-dem-profiles}), that the low-temperature enhancements in the $DEM$ profile are still present, even though heat transport in the lower atmosphere is still dominated by Coulomb collisions. Such enhancements relative to the predictions of a collision-dominated heat transport model \citep[cf.][]{2010ApJ...714.1290M} are indeed ubiquitously observed \citep[e.g.,][]{1981ApJ...247..686R}. This not only strongly indicates that turbulence plays a key role in thermal energy transport in active region loops, but also gives a probe to assess the value of the turbulence mean free path $\lambda_T$.

We have here demonstrated that the turbulent suppression of energy transport by thermal conduction, and the corresponding changes in the loop temperature profile, results in a $DEM$ profile that rises at low temperatures, broadly consistent with observations, and in much better agreement with observations than the $DEM$ profile produced by a model which incorporates collision-dominated conduction throughout. This explanation for the enhanced low-temperature $DEM$ is both simple and compelling, utilizes plasma conditions consistent with observations, and does not need to invoke external phenomena such as type II spicules \citep{2012JGRA..11712102K} and low-lying cool loops (although these phenomena may still, of course, contribute a component to the $DEM$ in some regions). Our results also suggest a powerful test for the presence of turbulence in coronal loops, in addition to observations of individual spectral line profiles, which directly reveal the presence of turbulence through excess Doppler broadening.

Based on the encouraging $DEM$ profiles obtained by incorporating turbulence-driven conduction in the upper levels of active region loops, we encourage the use of active region loop structures such as those derived here to model the surrounding atmosphere in models of energy release and transport in solar flares. We would also urge that such models incorporate a thermal conduction term that includes a level of turbulence consistent with that defining the background atmosphere.

The results of the present work are based on a turbulent mean free path $\lambda_T$ that is, following Equation~(17) of \cite{2016ApJ...824...78B}, assumed to be independent of velocity. However, it is also possible that the turbulent scattering mean free path {\it does} depend on velocity, such as the parametric power-law form $\lambda_T \propto v^\alpha$ used in Equation~(7) of \cite{2016ApJ...824...78B}. \cite{2018ApJ...865...67E} have studied the form of the heat conduction term for various power-law indices $\alpha$, and \cite{2022ApJ...931...60A} have further shown that observed profiles for the Fe XXI 1354~\AA\ spectral line observed by the Interface Region Imaging Spectrograph \citep[IRIS;][]{2014SoPh..289.2733D} are consistent with a value $\alpha=4$ (and a conduction suppression ratio $R = \lambda_C(v=v_{\rm thermal})/\lambda_T \simeq 1$). (Interestingly, for such a value of $\alpha$ the velocity dependencies of the collisional and turbulent mean free paths have identical $\lambda \propto v^4$ forms, so that the effect of introducing turbulence is simply to scale the mean free path downward: $1/\lambda = 1/\lambda_C + 1/\lambda_T = (1 + R)/\lambda_C$, so that $\lambda = \lambda_C/(1+R) = \lambda_C/2$.) Future modeling incorporating such a velocity-dependent $\lambda_T$, and comparison of the resulting differential emission measure profiles with those deduced from observations, is a clear next step to quantifying and parameterizing the role of turbulence in determining the thermodynamic structure of the active Sun.

\begin{acknowledgements}

We thank the referee for a very careful reading of the manuscript and for several useful and insightful comments, including one that allowed us to rectify a rather significant error in an earlier version of the manuscript. AGE was supported by NASA Kentucky under NASA award number 80NSSC21M0362; SJB was supported by NSF CAREER award AGS-1450230.

\end{acknowledgements}

\bibliographystyle{aasjournal}
\bibliography{bib-scaling-laws}

\end{document}